\documentclass[a4paper,11pt]{article}
\pdfoutput=1 

\usepackage{jinstpub} 
\usepackage{lineno}

\usepackage{caption}
\usepackage{subcaption}
\usepackage{siunitx}
\usepackage[percent]{overpic}

\title{GiBUU based neutrino interaction simulations in KM3NeT}

\author[a]{J. Schumann,\note{Corresponding author.}}
\author[b]{B. Jung}

\affiliation[a,1]{Friedrich-Alexander-Universit\"at Erlangen-N\"urnberg, Erlangen Centre for Astroparticle Physics, Erwin-Rommel-Straße
1, 91058 Erlangen, Germany}
\affiliation[b]{Nikhef, National Institute for Subatomic Physics, PO Box 41882, Amsterdam, 1009 DB Netherlands}
\emailAdd{johannes.schumann@fau.de}

\abstract{
The simulation of the neutrino interaction is a crucial step in the simulation chain
of a neutrino experiment. The different processes taking part in the neutrino scattering 
on a nucleus require several approximations in order to make the simulation possible 
and to realize reasonable computation times.
This can be realised in different ways, e.g. by parametrised models for the different scattering processes and energy regimes as it is implemented in GENIE.
The GiBUU neutrino generator utilises the Boltzmann-Uehling-Uhlenbeck equation to 
simulate the particle flow after the neutrino interactions, the so-called final state 
interactions. 
The detector-specific results in form of the visible energy in the detector after
the light propagation simulation and the KM3NeT event reconstruction are presented.
In addition to that, the comparison to the GENIE based simulation environment in KM3NeT (gSeaGen) is drawn.}


\collaboration[c]{on behalf of the KM3NeT collaboration}

\proceeding{VLVnT 2021 - Very Large Volume Neutrino Telescope Workshop\\
  18-21 May 2021\\
  Valencia (Online)}

\notoc
\begin{document}
\maketitle
\flushbottom

\section{Introduction}
\label{sec:intro}
The KM3NeT neutrino telescope is currently being built in the depths of the Mediterranean Sea. 
It will host two water Cherenkov detectors comprising 3D-arrays of photomultiplier tubes (PMT) 
with different layouts. 
The detector designed for the low energy regime is called ORCA (Oscillation Research with Cosmics in the Abyss) in order to tackle the determination
of the neutrino mass hierarchy. In this lower energy regime the secondary particles from the neutrino interaction 
are distributed more isotropically and deviate further from the direction 
of the primary neutrino. 
Also the reconstruction of the energy depends stronger on energy thresholds for the
emission of the secondary particles, thus information about the interaction is 
crucial in order to gain knowledge about the light distribution of low-energy 
event topologies and subsequently improve the energy and direction reconstruction of the primary neutrino \cite{2021:aiello:DeterminingNeutrinoMass}.
\section{Neutrino Interaction Simulation}
In order to make a numerical simulation of
neutrino scattering processes with so-called neutrino generators possible
within reasonable computation times several approximations are necessary.  
The neutrino generators GiBUU and GENIE implement the neutrino interaction via 
a factorised model, i.e. splitting the scattering event into the initial electroweak 
scattering and the propagation of the generated particles 
through the nucleus \cite{2012:buss:TransporttheoreticalDescriptionNuclear, 2015:andreopoulos:GENIENeutrinoMonte}. In GiBUU this 
propagation of final states is done via propagating phase space densities utilising the
Boltzmann-Uehling-Uhlenbeck equation. In order to adapt the simulated neutrino 
interactions to the KM3NeT environment, the neutrino generators are wrapped 
by custom software packages. For GENIE, the KM3NeT collaboration has developed a wrapper 
called gSeaGen \cite{2020:aiello:GSeaGenKM3NeTGENIEbased} and for GiBUU  
the KM3BUU software package (see Section \ref{sec:km3buu}) has been developed.
\subsection{Final State Particle Picture}
\label{sec:intsim:subsec:fsi}
The resulting particles after the propagation through the nucleus are 
written out by the generator and are referred to as final state particles.
The contribution of the different particle types over energy is compared via
the production cross section $\frac{d\sigma_X}{dE_\nu}$,
where $X$ stands for the individual secondary particle type.
\begin{figure}[!htbp]
    \centering
    \begin{overpic}[width=1.1\textwidth, trim=0 399 0 100, clip]{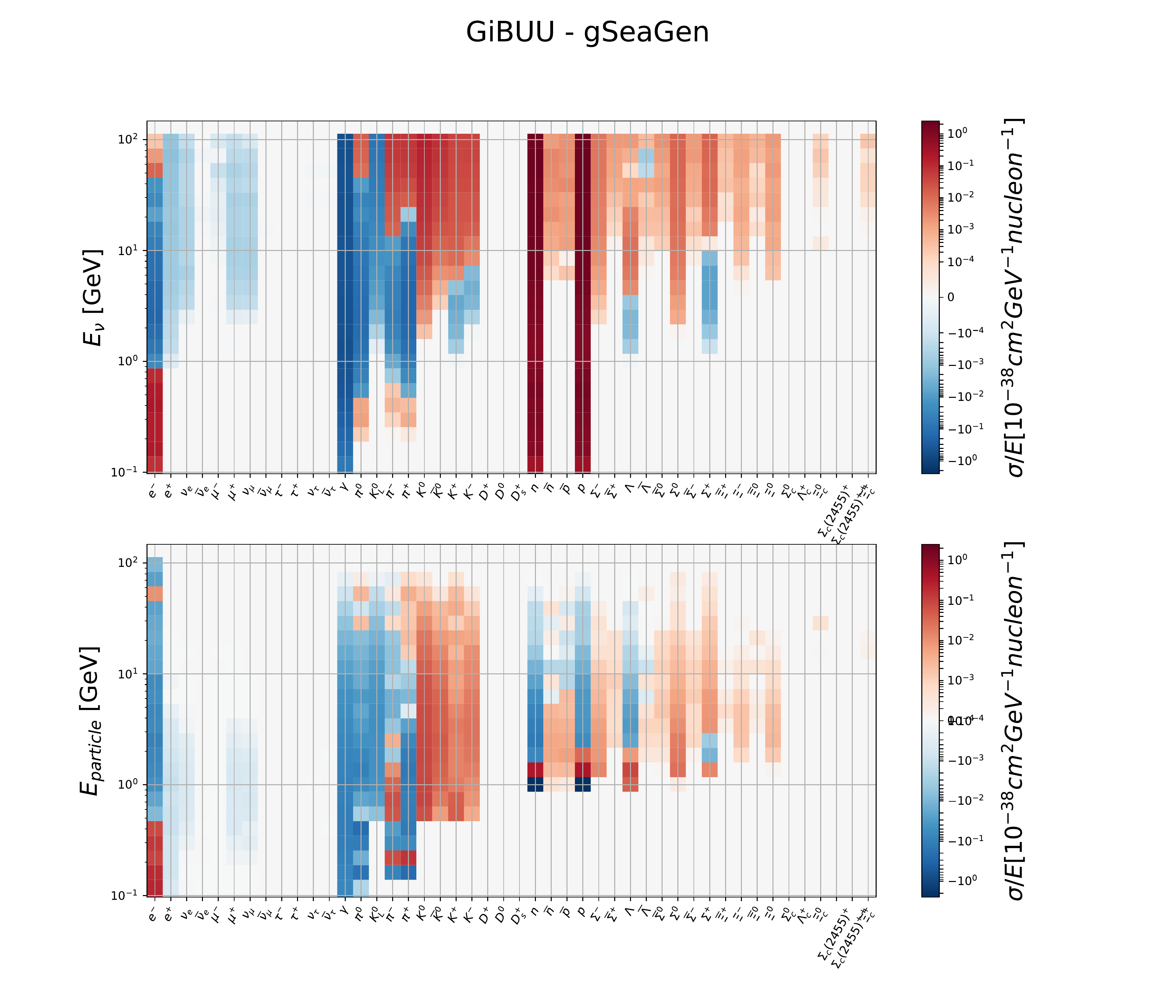}
        \put (53,10) {\footnotesize KM3NeT Preliminary}
    \end{overpic}
    \caption{Difference of the production cross sections, i.e. the GENIE cross section subtracted from the GiBUU cross section: $\left.\frac{d\sigma_X}{dE_\nu}\right|_\text{GiBUU}-\left.\frac{d\sigma_X}{dE_\nu}\right|_\text{GENIE}$.}
    \label{sec:intsim:subsec:fsi:fig:diff}
\end{figure}
In the regime below \SI{10}{\giga\eV} the final state particle types 
consist 
of photons and\\ $\pi$-mesons (besides the out going lepton from the neutrino
vertex itself) and the emission of secondary particles from GENIE is higher (see Figure \ref{sec:intsim:subsec:fsi:fig:diff}). 
At higher energies also heavier mesons, i.e. Kaons, are generated and their
production cross section from GiBUU data exceeds the one from GENIE.
%% Referee change
The GiBUU dataset does not contain photons, which yields the dominant contribution
of photons for the GENIE dataset. The difference in the production cross section
for protons and neutrons is currently under further investigation. The mean energy 
of those secondary nucleons is $\bar{E}_{p,n} = \SI{1.08}{\giga\eV}$ 
($\sigma_{p,n}=\SI{0.99}{\giga\eV}$) which is close to the rest energy of the particles.
Thus, the contribution to the visible light yield in the detector is expected to
be neglectable.
%%%
\subsection{Visible Energy}
\begin{figure}[!htbp]
    \centering
    \begin{subfigure}{0.37\textwidth}
        \begin{overpic}[width=\textwidth, trim=0 0 0 40, clip]{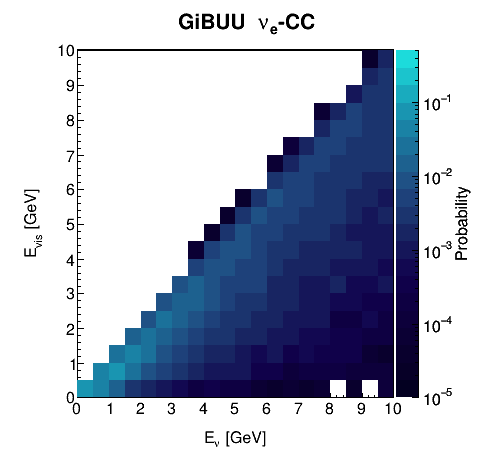}
            \put (20,80) {\tiny KM3NeT Preliminary}
        \end{overpic}
    \end{subfigure}
    \begin{subfigure}{0.5\textwidth}
        \begin{overpic}[width=\textwidth, trim=0 15 85 50, clip]{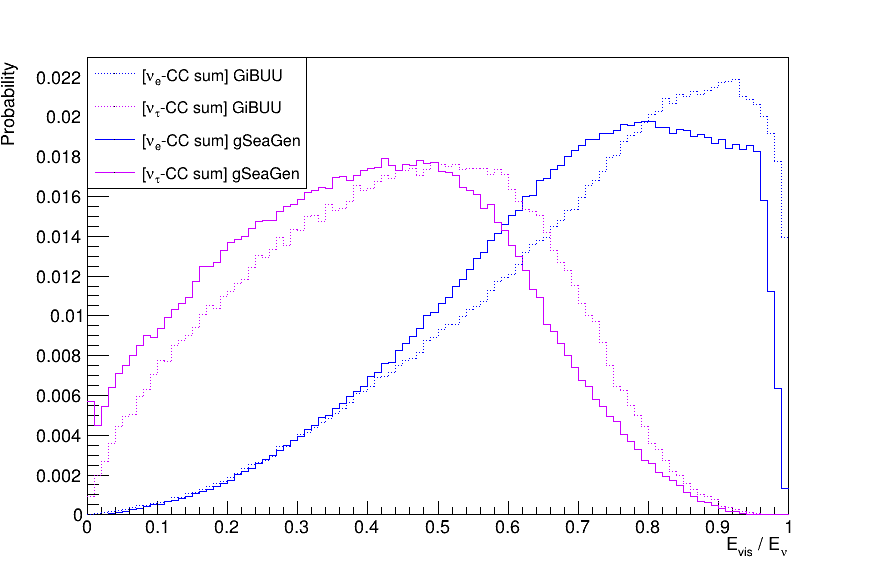}
            \put (40,58) {\tiny KM3NeT Preliminary}
        \end{overpic}
    \end{subfigure}
    \caption{Visible energy $E_\text{vis}$ over the primary neutrino energy $E_\nu$ for GiBUU (left) and the ratio $E_\text{vis}/E_\nu$ for $\nu_{e,\tau}$-CC-interactions for GiBUU and gSeaGen in the energy range $E_\nu\in[\SI{1}{\giga\eV},\SI{10}{\giga\eV}]$ (right).}
    \label{sec:vise:fig:evisratio}
\end{figure}

The visible energy of a given final state is defined as the energy
contained in an electromagnetic shower which is converted to photons.
The definition was applied and the values for the parametrisation of the used rational 
function are taken from \cite{2012:dentler:InvestigationOneParticleApproximation}.
The visible energy distribution for both generators is shown in Figure \ref{sec:vise:fig:evisratio}.
In the used configuration the distribution of the ratio $E_\text{vis}/E_\nu$ for the KM3BUU data has a slight shift to higher values. This seems to contradict the results from Section \ref{sec:intsim:subsec:fsi}, but 
those are given in units of production cross section which is independent from the 
simulated flux $\Phi(E)\approx E^{-1}$. 
\subsection{Event Trigger \& Reconstruction}
\label{sec:intsim:subsec:reco}
\begin{figure}
    \centering
    \begin{subfigure}[b]{0.43\textwidth}
        \begin{overpic}[width=\textwidth]{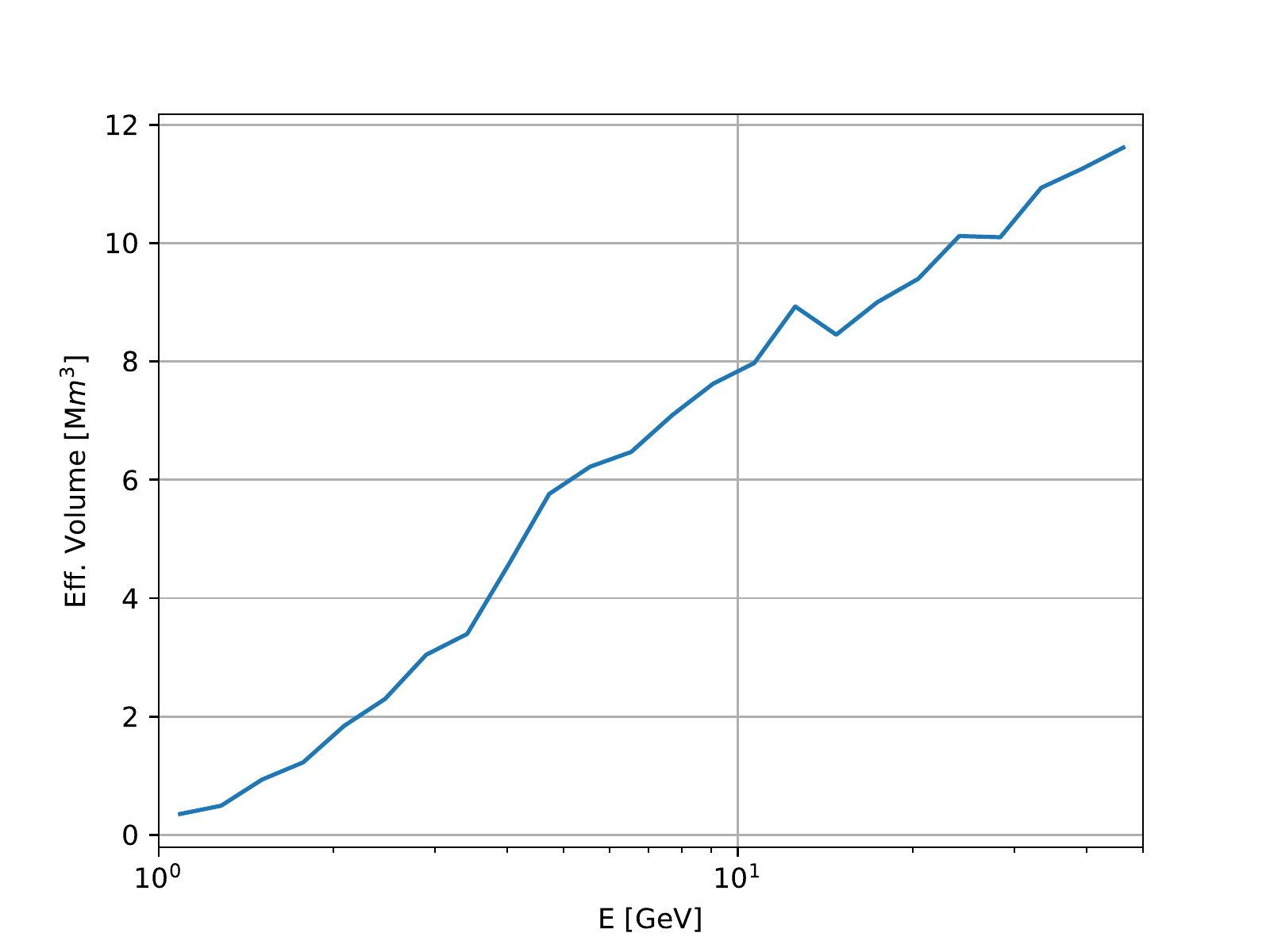}
            \put (14, 60) {\footnotesize KM3NeT Preliminary}
        \end{overpic}
    \end{subfigure}
    \begin{subfigure}[b]{0.42\textwidth}
        \includegraphics[width=\textwidth]{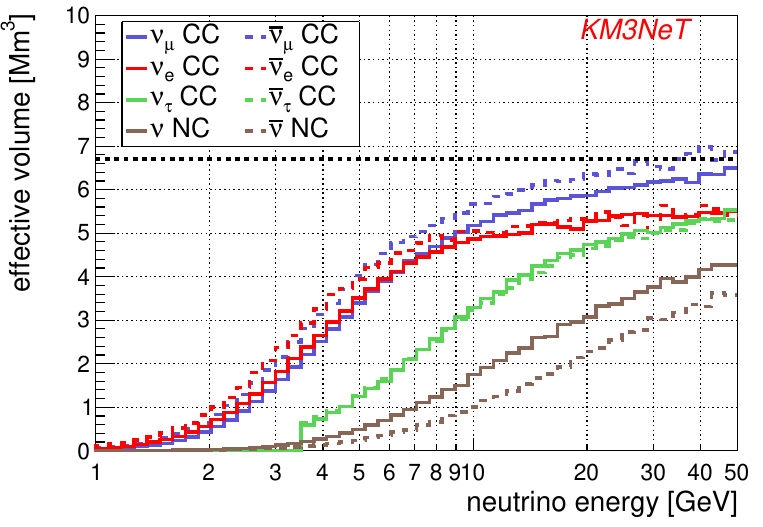}
    \end{subfigure}
    \caption{Effective volume for the reconstructed events for the GiBUU dataset (left)
    and offical estimate for KM3NeT taken from \cite{2021:aiello:DeterminingNeutrinoMass} (right).}
    \label{sec:intsim:subsec:rec:fig:effvol}
\end{figure}

\begin{figure}[!htbp]
    \centering
    \begin{subfigure}[b]{0.46\textwidth}
        \begin{overpic}[width=\textwidth]{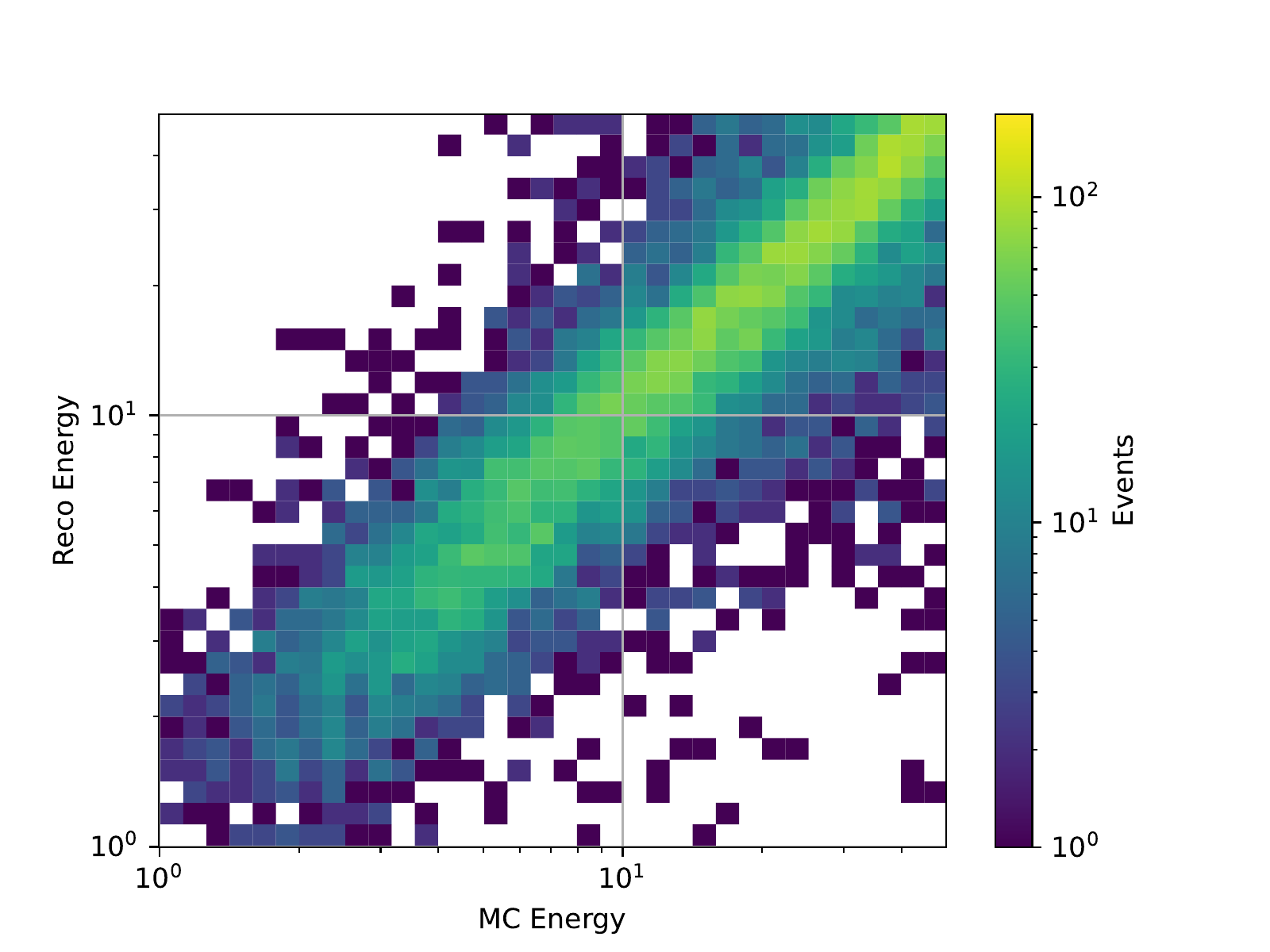}
            \put (14, 62) {\tiny KM3NeT Preliminary}
        \end{overpic}
    \end{subfigure}
    \begin{subfigure}[b]{0.46\textwidth}
        \begin{overpic}[width=\textwidth]{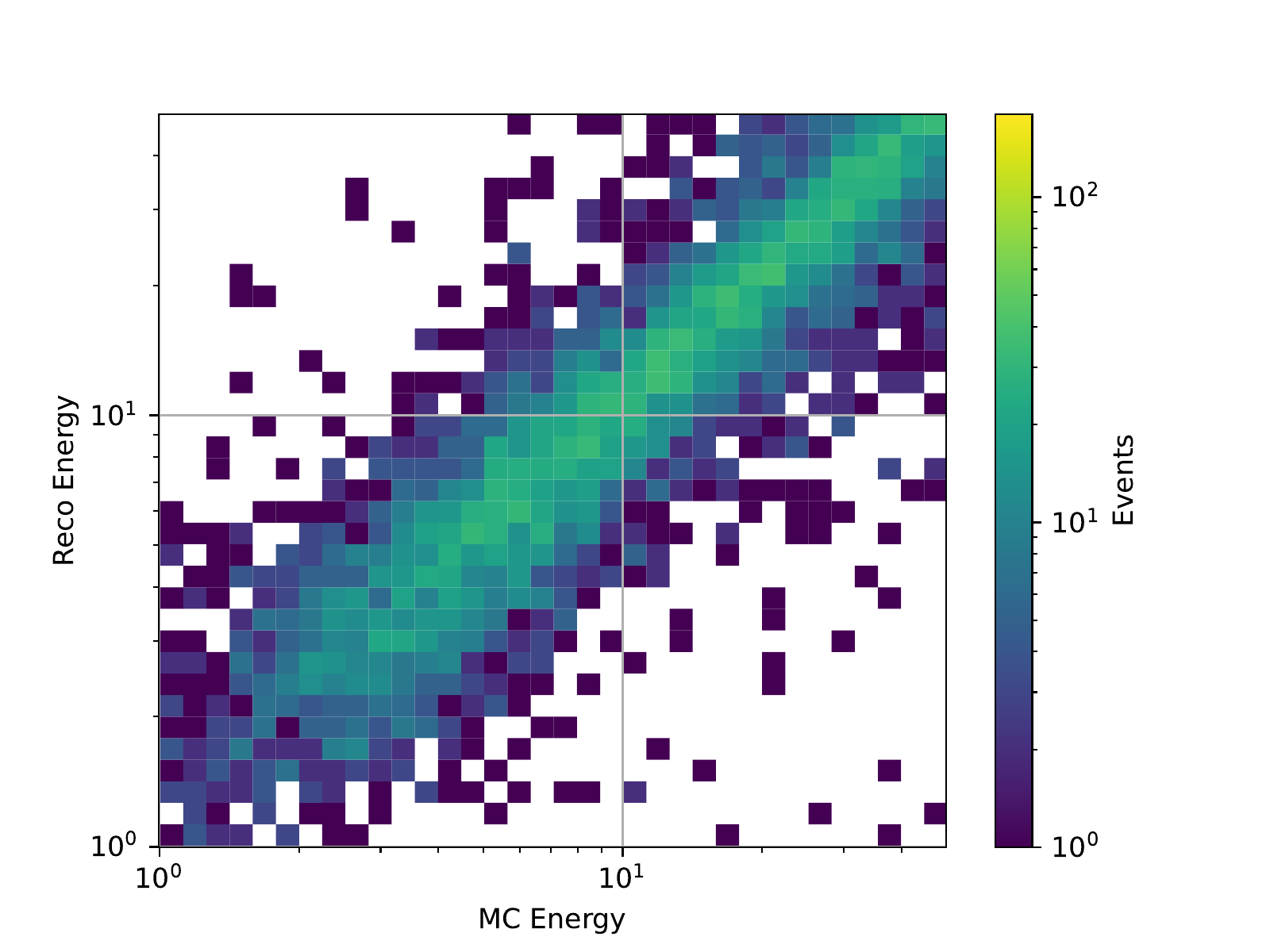}
            \put (14, 62) {\tiny KM3NeT Preliminary}
        \end{overpic}
    \end{subfigure}
    \caption{Comparison of the reconstructed event shower energy and the simulated 
             (MC) energy of GiBUU (left) and gSeaGen (right) normalised to a uniform 
             distribution of neutrino events in the used energy range.}
    \label{sec:intsim:subsec:reco:fig:reco}
\end{figure}

In order to get a realistic detector picture including the environment  
properties and geometry effects, in the next step the neutrino 
generator output is fed into a light generation and propagation simulation
resulting in photosensor readings (hits). With the hit information the simulation 
data contains the same shape compared to the 
DAQ data from the detector and can subsequently be fed into the subsequent 
data processing tools, i.e. triggering and reconstruction.
The normalisation of the number of events over energy can be compared via the
effective volume $V_\text{eff}$ defined as $V_\text{gen} \cdot {N_\text{det}}/{N_\text{gen}}$, 
which is shown in Figure \ref{sec:intsim:subsec:rec:fig:effvol}. 
It stands out, that the distribution for GiBUU follows the expected distribution 
in the energy regime between \SI{1}{\giga\eV} and \SI{10}{\giga\eV}, but exceeds
the expected upper limit at \SI{7}{\mega\metre\cubed} and continues rising at 
energies higher than \SI{10}{\giga\eV}. This deviation from the expectation
is currently further investigated.
In Figure \ref{sec:intsim:subsec:reco:fig:reco} the comparison 
between KM3BUU and gSeaGen is shown for the reconstructed energy over the true Monte-Carlo energy.
The ridge of the events lies for both on the expected straight line of
$E_\text{reco}/E_\text{MC} = 1$, which indicates similar magnitude of systematic effects 
from the processed KM3BUU dataset compared to the gSeaGen dataset.
The origin of systematics effects have not been studied yet.
The distributions yield a lower reconstruction 
efficiency for gSeaGen at higher energies above \SI{10}{\giga\eV} compared to KM3BUU.
This difference can be explained by the normalisation comparing the excess of 
effective volume above \SI{10}{\giga\eV} for KM3BUU (see Figure \ref{sec:intsim:subsec:rec:fig:effvol}).
\section{KM3BUU}
\label{sec:km3buu}
The KM3BUU framework is a python based wrapper for the GiBUU Fortran environment,
which is installed and operated inside a singularity container. 
The general GiBUU workflow for running the simulation is unchanged, but can be completely
steered via the KM3BUU python environment. The GiBUU output consists of multiple
files which carry the full information about the simulated neutrino scatterings.
The output directory containing those GiBUU output files can be 
parsed with the KM3BUU framework and converted to multiple data formats, e.g.
awkward arrays \cite{2021:pivarski:ScikithepAwkward015} and pandas dataframes \cite{2021:reback:PandasdevPandasPandas} for the pure neutrino generator output or
converted to the KM3NeT dataformat. 
This KM3NeT dataformat is then completely compatible to the KM3NeT (simulation) tools . 
\section{Conclusion}
The KM3BUU package is a quite new software package and the technical aspect of 
software errors is non-neglectable, but the distributions along the simulation steps 
do not show any major discrepancies compared to GENIE.
The difference in the $\pi$-meson production rate in the lower \si{\giga\eV} regime 
is expected to be higher compared to previous studies, e.g. Coloma et al \cite{2014:coloma:NeutrinonucleusInteractionModels}, but this depends on the GENIE input cross section and 
the simulation configuration of GiBUU.
This difference affects also the event topology starting in the analysis step
of the visible energy.
In the final step the event reconstruction shows no systematic offsets or deviations 
with focus on the reconstructed energy. The excess of effective volume visible for
the KM3BUU data follows the trend of a higher visible energy fraction $R$, but 
as this exceeds the estimate of previous studies significantly this is currently 
further investigated.
\bibliographystyle{JHEP}
\bibliography{references}
\end{document}